\documentclass{article}
\begin{document}
\title{Relativity of $\hbar$}
\author{Jos\'e M. Isidro \\ 
Department of Theoretical Physics,\\ 
1 Keble Road, Oxford OX1 3NP, UK\\ 
and\\
Instituto de F\'{\i}sica Corpuscular (CSIC-UVEG)\\
Apartado de Correos 22085, 46071 Valencia, Spain\\
{\tt isidro@thphys.ox.ac.uk}}
\maketitle

\begin{abstract}

Looking for a quantum--mechanical implementation of duality,
we formulate a relation between coherent states and complex--differentiable 
structures on classical phase space ${\cal C}$. 
A necessary and sufficient condition for the existence of 
locally--defined coherent states is the existence of an almost complex 
structure on ${\cal C}$. A necessary and sufficient condition 
for globally--defined coherent states is a complex structure on ${\cal C}$.  

The picture of quantum mechanics that emerges is conceptually close to that of a 
geometric manifold covered by local coordinate charts. Instead of the 
latter, quantum mechanics has local coherent states. A change of 
coordinates on ${\cal C}$ may or may not be holomorphic. Correspondingly, 
a transformation between quantum--mechanical states may or may not preserve 
coherence. Those that do not preserve coherence are duality 
transformations. A duality appears as the possibility of giving 
two or more, apparently different, descriptions 
of the same quantum--mechanical phenomenon.
Coherence becomes a local property on classical phase space. 
Observers on ${\cal C}$ not connected by means of a holomorphic change of 
coordinates need not, and in general will not, agree on what is a semiclassical 
effect {\it vs.} what is a strong quantum effect. 

Keywords: Coherent states, almost complex manifolds, duality.

2001 Pacs codes: 03.65.Bz, 03.65.Ca, 03.65.-w.

\end{abstract}


\tableofcontents

\section{Introduction}\label{intro}

\subsection{Motivation}\label{moti}

Coherent states are quantum--mechanical states that enjoy semiclassical 
properties \cite{COHST}. One definition of coherent states uses the Heisenberg 
inequality $\Delta q\Delta p \geq \hbar/2$. The latter is saturated precisely 
by coherent states. Planck's constant $\hbar$ can be interpreted 
as a parameter measuring how far quantum mechanics deviates from classical mechanics. 

Along a different line of developments, the notion of duality plays a key role 
in recent breakthroughs in the quantum theories of fields, strings and branes 
\cite{DUALITY, LAG, VAFA}. Broadly speaking, under duality one understands a transformation 
of a given theory, in a certain regime of the variables and parameters that define it, 
into a physically equivalent theory with different variables and parameters. 
Often, what appears to be a highly nontrivial quantum excitation in a given theory 
turns out to be a simple perturbative correction from the viewpoint 
of a theory dual to the original one. This suggests that what constitutes 
a quantum correction may be a matter of convention: the notion of classical 
{\it vs.} quantum is relative to which theory the measurement is made from.  
In this way we arrive at the conclusion that may $\hbar$ depend on the observer. 
This, in turn, implies that coherence is also theory--dependent, or observer--depedent.
What one observer calls coherent need not be coherent to another observer.

The standard formulation of quantum mechanics \cite{GALINDO}
does not allow for such a relativity of the concept of a quantum. 
Somewhat imprecisely, we will call this concept {\it relativity of $\hbar$}. 
The limit $\hbar\to 0$ is the {\it  semiclassical regime}, and the limit $\hbar\to\infty$  
is the {\it  strong quantum regime}. Under the usual formulation of quantum mechanics, 
if one observer calls a certain phenomenon {\it semiclassical}, 
then so will it be for all other observers. If one observer calls 
a certain phenomenon {\it strong quantum}, then so will it be for all other observers.

In view of these developments, a framework for quantum mechanics 
is required that can accommodate such a relativity of $\hbar$ \cite{VAFA}.
Generalising the geometric approach of ref. \cite{ASHTEKAR}, 
in ref. \cite{ME} we have explicitly developed one such framework. 
There remains the alternative, though equivalent, possibility of addressing 
the relativity of $\hbar$ through an analysis of coherent states.
Other geometric approaches to quantum mechanics dealing with this and 
related topics are refs. 
\cite{ANANDAN, MATONE, FH, GEOMKLAUDER, MILANO, TIWARI, MEX}.

\subsection{Summary of main results}\label{suma}

The purpose of this article is to present a framework for quantum mechanics in which
coherent states are defined locally on classical phase space ${\cal C}$, 
but not necessarily globally. This is an explicit implementation of a relativity for $\hbar$.
Globally--defined coherent states are the rule in the standard 
presentations of quantum mechanics. As such they preclude any possible 
observer--dependence for $\hbar$.

We will analyse the relationship between complex--differentiable structures
on ${\cal C}$ and coherent states of the corresponding quantum mechanics. 
We will elaborate on a suggestion presented in ref. \cite{ME}, according to which coherent 
states may not be globally defined for all observers on ${\cal C}$, 
and we will relate this observation to the (in)existence of complex structures on ${\cal C}$. 

To our purposes it is enough to realise that there exist finite--dimensional symplectic 
manifolds ${\cal C}$ admitting no complex structure \cite{MCDUFF}. 
Using the symplectic structure, we will construct local coherent states around a certain point 
on ${\cal C}$. Upon a nonholomorphic change of Darboux coordinates to another point on 
${\cal C}$, those states will cease to be coherent. 

Our results may be summarised as follows. Coherent states can always 
be defined locally, {\it i.e.}, in the neighbourhood of any point  on ${\cal C}$. 
This is merely a restatement, in physical terms, of Darboux's theorem for symplectic 
manifolds \cite{ARNOLD}. When there is a complex 
structure ${\cal J}_{\cal C}$, coherence becomes a global property on ${\cal C}$. 
In the absence of a complex structure, however, the best we can do 
is to combine Darboux coordinates $q$, $p$ as $q+{\rm i}p$. 
Technically this only defines an almost complex structure $J_{\cal C}$ on ${\cal C}$
\cite{KN}. Since the combination $q+{\rm i}p$ 
falls short of defining a complex structure, quantities depending on $q+{\rm i}p$ 
on a certain coordinate patch will generally also depend on $q-{\rm i}p$ 
when transformed to another coordinate patch. This proves that coherence remains 
a local property on classical phase space: observers not connected by means of 
a holomorphic change of coordinates need not, and in general will not,
agree on what is a semiclassical effect {\it vs.} what is a strong quantum effect. 

\subsection{Outline}\label{outline}

This article is organised as follows. Section \ref{clqu} sets the scene 
with a quick review of the geometry of classical and quantum phase spaces, 
denoted  ${\cal C}$ and ${\cal Q}$ respectively. The construction of coherent 
states from a complex structure ${\cal J}_{\cal C}$ on ${\cal C}$ is recalled 
in section \ref{globcoh}; further issues discussed in this section are the 
uniqueness of the vacuum and the global character of the coherent states so constructed. 
In section \ref{loccoh} we relax the complex structure ${\cal J}_{\cal C}$ 
to an almost complex structure $J_{\cal C}$. In so doing we observe that the 
vacuum state is only locally defined on ${\cal C}$ and that, under a nonholomorphic 
change of coordinates on ${\cal C}$, global coherence is lost. In section \ref{prov} 
we prove the coherence of the global states constructed in section \ref{globcoh}, 
and the coherence of the local states constructed in section \ref{loccoh}.

The conditions under which a complex structure $J_{\cal C}$ lifts to a complex 
structure ${\cal J}_{\cal C}$ are known in the mathematical literature; 
we recast them in physical terms in section \ref{when}. In section \ref{dual} 
we introduce the concept of a holomorphic foliation ${\cal F}$ of ${\cal C}$. 
The leaves of ${\cal F}$ are the maximal holomorphic submanifolds of ${\cal C}$ 
on which coherence remains a globally--defined property. Nonholomorphic coordinate 
changes between different holomorphic leaves of ${\cal F}$ are interpreted as duality 
transformations of the quantum theory on ${\cal Q}$. Some geometric examples serve 
to illustrate our proposal. Finally, section \ref{discussion} 
discusses some physical and mathematical aspects of our construction.

\section{Classical and quantum phase spaces}\label{clqu}

Our starting point is an infinite--dimensional, complex, separable Hilbert space 
of quantum states, ${\cal H}$, that is most conveniently viewed as a real vector 
space equipped with a complex structure ${\cal J}$. Correspondingly, 
the Hermitian inner product can be decomposed into real and imaginary parts,
\begin{equation}
\langle\phi,\psi\rangle = g(\phi,\psi) + {\rm i}\omega(\phi,\psi),
\label{dec}
\end{equation}
with $g$ a positive--definite, real scalar product and $\omega$ a symplectic form.
The metric $g$, the symplectic form $\omega$ and the complex structure 
${\cal J}$ are related as
\begin{equation}
g(\phi,\psi)=\omega(\phi,{\cal J}\psi),
\label{kah}
\end{equation}
which means that the triple $({\cal J}, g, \omega)$ endows the Hilbert space 
${\cal H}$ with the structure of a K\"ahler space \cite{KN}. 

Let ${\cal Q}$ denote the  space of unit rays in ${\cal H}$, and let $\omega_{\cal Q}$ 
be the restriction to ${\cal Q}$ of the symplectic form $\omega$ on ${\cal H}$. 
The space ${\cal Q}$, called {\it quantum phase space}, is an  infinite--dimensional
symplectic manifold. 
Classical phase space ${\cal C}$ is a $2n$--dimensional symplectic 
manifold; let $\omega_{\cal C}$ denote its symplectic form. If $q^l$, $p_l$, $l=1,\ldots, n$, 
are local Darboux coordinates on ${\cal C}$, we have
\begin{equation}
\omega_{\cal C}=\sum_{l=1}^n{\rm d} p_l \wedge {\rm d} q^l.
\label{cano}
\end{equation}

It should be observed that, while both ${\cal C}$ and ${\cal Q}$ 
are symplectic manifolds, the latter is always K\"ahler, while the former 
need not be K\"ahler. However, in the framework of geometric \cite{GQUANT} 
and deformation \cite{BEREZIN, SCHLICHENMAIER, GARCIA} quantisation it is customary 
to consider the case when ${\cal C}$ is a compact K\"ahler manifold. 
In this context one introduces the notion of a quantisable, compact, K\"ahler 
phase space ${\cal C}$. This means that there exists an associated quantum 
line bundle $(L, g, \nabla)$ on ${\cal C}$, where $L$ is a holomorphic line 
bundle, $g$ a Hermitian metric on $L$, and $\nabla$ a connection 
compatible with the complex structure and the Hermitian metric.
Furthermore, the curvature $F$ of the connection $\nabla$ and the K\"ahler form 
$\omega_{\cal C}$ are required to satisfy
\begin{equation}
F=-{\rm i} \omega_{\cal C}.
\label{quantti}
\end{equation}
It turns out that quantisable, compact K\"ahler manifolds are projective 
algebraic manifolds and viceversa \cite{SCHLICHENMAIER}. For the purpose 
of introducing duality transformations, however, the previous assumptions 
are too restrictive. At most we will require ${\cal C}$ to support a complex 
structure ${\cal J}_{\cal C}$ compatible with the symplectic structure 
$\omega_{\cal C}$, as in the next section.

\section{Global coherent states from a complex structure}\label{globcoh}

Let us assume that ${\cal C}$ admits a complex structure ${\cal J}_{\cal C}$. 
Furthermore let ${\cal J}_{\cal C}$ be compatible with the symplectic structure 
$\omega_{\cal C}$. This means that the real and imaginary parts of the holomorphic 
coordinates $z^l$ for ${\cal J}_{\cal C}$ are Darboux coordinates for the symplectic form 
$\omega_{\cal C}$:
\begin{equation}
z^l=q^l + {\rm i} p_l,  \qquad l=1,\ldots, n.
\label{compcoords}
\end{equation}
The set of all $z^l$ so defined provides a holomorphic atlas for ${\cal C}$.
Upon quantisation, the Darboux coordinates
$q^l$ and $p_l$ become operators $Q^l$ and $P_l$ on ${\cal H}$ satisfying the Heisenberg algebra 
\begin{equation}
[Q^j, P_k]={\rm i}\delta^j_k.
\label{hei}
\end{equation}
Define the annihilation operators
\begin{equation}
A^l=Q^l+{\rm i} P_l,\qquad l=1,\ldots, n.
\label{anni}
\end{equation}
Quantum excitations are measured with respect to a vacuum state 
$|0\rangle$. The latter is defined as that state in ${\cal H}$ which
satisfies
\begin{equation}
A^l|0\rangle = 0,  \qquad l=1, \ldots, n,
\label{vac}
\end{equation}
and coherent states $|z^l\rangle$ are eigenvectors of $A^l$, 
with eigenvalues given in equation (\ref{compcoords}) above:
\begin{equation}
A^l|z^l\rangle=z^l|z^l\rangle,\qquad l=1,\ldots, n.
\label{annop}
\end{equation}

How do the vacuum state $|0\rangle$ and the coherent states $|z^l\rangle$
transform under a canonical coordinate transformation on ${\cal C}$? 
Call the new coordinates $q'^l$, $p'_l$. As they  are Darboux, they continue 
to satisfy equation (\ref{cano}). Upon quantisation the 
corresponding operators $Q'^l$, $P'_l$ continue to satisfy the Heisenberg 
algebra (\ref{hei}). Then the combinations
\begin{equation}
z'^l=q'^l + {\rm i} p'_l,  \qquad l=1,\ldots, n
\label{xcompcoords}
\end{equation}
continue to provide holomorphic coordinates for ${\cal C}$, and the 
transformation between the $z^l$ and the $z'^l$ is given by an $n$--variable
holomorphic function $f$,
\begin{equation}
z'=f(z),\qquad \bar\partial f=0.
\label{fhol}
\end{equation}
We can write as above
\begin{equation}
A'^l=Q'^l+{\rm i} P'_l,\qquad l=1,\ldots, n,
\label{annix}
\end{equation}
\begin{equation}
A'^l|0\rangle = 0,  \qquad l=1, \ldots, n,
\label{vacx}
\end{equation}
\begin{equation}
A'^l|z'^l\rangle=z'^l|z'^l\rangle,\qquad l=1,\ldots, n.
\label{annopx}
\end{equation}
There is no physical difference between equations (\ref{anni}), (\ref{vac}) and 
(\ref{annop}), on the one hand, and their holomorphic transforms (\ref{annix}), 
(\ref{vacx}) and (\ref{annopx}), on the other. Under the transformation 
(\ref{fhol}), the vacuum state $|0\rangle$ is mapped into itself, 
and the coherent states $|z^l\rangle$ are mapped into the coherent states $|z'^l\rangle$.
Therefore the notion of coherence is global for all observers on ${\cal 
C}$, {\it i.e.}, any two observers will agree on what is a coherent 
state {\it vs.} what is a noncoherent state. A consequence of 
this fact is the following. Under symplectomorphisms (or, equivalently, 
holomorphic diffeomorphisms) of ${\cal C}$, the semiclassical regime 
of the quantum theory on ${\cal Q}$ is mapped into the semiclassical regime, 
and the strong quantum regime is mapped into the strong quantum regime.

Conversely, one can reverse the order of arguments in this section. 
Start from the assumption that one can define global coherent states 
$|z^l\rangle$ and a global vacuum $|0\rangle$ on the symplectic manifold ${\cal C}$.
{\it Globality}\/ here does not mean that one can cover all of ${\cal C}$ 
with just one coordinate chart (which is impossible if ${\cal C}$ is compact).
Rather it means that, under all
symplectomorphisms of ${\cal C}$, the vacuum is mapped into itself, 
and coherent states are always mapped into coherent states. 
Then the coordinates $z^l$ defined by the eigenvalue equations 
(\ref{annop}) provide a local chart for ${\cal C}$. Collecting together the set of all 
such possible local charts we obtain an atlas for ${\cal C}$. This atlas is holomorphic 
thanks to the property of globality.

To summarise, the existence of a complex structure ${\cal J}_{\cal C}$ 
is equivalent to the existence of a globally defined vacuum and globally 
defined coherent states.

\section{Local coherent states from an almost complex structure}\label{loccoh}

We now relax the conditions imposed on ${\cal C}$. In this section we will assume
that ${\cal C}$ carries an almost complex structure $J_{\cal C}$ compatible with 
the symplectic structure $\omega_{\cal C}$. (When ${\cal C}$ is compact, 
the symplectic structure $\omega_{\cal C}$ automatically leads to an almost complex 
structure $J_{\cal C}$, so the assumption of the existence of a $J_{\cal C}$
compatible with $\omega_{\cal C}$ is in fact unnecessary; in that case, the 
symplectic structure $\omega_{\cal C}$ suffices \cite{MCDUFF}). 

Specificallly, an almost complex structure is defined  
as a tensor field $J_{\cal C}$ of type $(1,1)$ such that, at every point of 
${\cal C}$, $J_{\cal C}^2=-{\bf 1}$ \cite{KN}.  Using Darboux coordinates $q^l$, 
$p_l$ on ${\cal C}$ let us form the combinations
\begin{equation}
w^l=q^l + {\rm i} p_l, \qquad l=1,\ldots, n.
\label{combw}
\end{equation} 
Compatibility between $\omega_{\cal C}$ and $J_{\cal C}$ means that we can take $J_{\cal C}$ 
to be 
\begin{equation}
J_{\cal C}\left({\partial\over\partial w^l}\right)={\rm i}{\partial\over\partial w^l},
\qquad 
J_{\cal C}\left({\partial\over\partial \bar w^l}\right)=-{\rm i}{\partial\over\partial\bar w^l}.
\label{jota}
\end{equation}
Unless ${\cal C}$ is a complex  manifold to begin with, equations 
(\ref{combw}) and (\ref{jota}) fall short of defining a complex structure ${\cal J}_{\cal C}$. 
The set of all such $w^l$ does not provide a holomorphic atlas for ${\cal C}$. 
There exists at least one canonical coordinate transformation between 
Darboux coordinates, call them $(q^l, p_l)$ and $(q'^l, p'_l)$, such that 
the passage between $w^l=q^l+{\rm i}p_l$ and $w'^l=q'^l+{\rm i}p'_l$
is given by a nonholomorphic function $g$ in $n$ variables,
\begin{equation}
w'=g(w,\bar w), \qquad \bar\partial g\neq 0.
\label{nonhol}
\end{equation}

Mathematically, nonholomorphicity implies the mixing of $w^l$ and $\bar w^l$.
Quan-tum--mechanically, the loss of holomorphicity has deep physical consequences.
One would write, in the initial coordinates $w^l$, a defining equation for the vacuum 
state $|0\rangle$
\begin{equation}
a^l|0\rangle =0, \qquad l=1,\ldots, n,
\label{xvac}
\end{equation}
where $a^l=Q^l + {\rm i} P_l$ is the corresponding local annihilation operator.
However, one is just as well entitled to use the new coordinates $w'^l$
and write
\begin{equation}
a'^l|0'\rangle =0, \qquad l=1,\ldots, n,
\label{xvacu}
\end{equation}
where we have primed the new vacuum, $|0'\rangle$. 
Are we allowed to identify the states $|0\rangle$ and $|0'\rangle$? 
We could identify them if $w'^l$ were a holomorphic function of $w^l$; 
such was the case in section \ref{globcoh}. However, now we are considering 
a nonholomorphic transformation, and we cannot remove the 
prime from the state $|0'\rangle$. This is readily proved.
We have
\begin{equation}
a'=G(a, a^{\dagger}),
\label{effe}
\end{equation}
with $G$ a quantum nonholomorphic function corresponding to the classical 
nonholomorphic function $g$ of equation (\ref{nonhol}). As $[a^j,a_k^{\dagger}]=\delta^j_k$, 
ordering ambiguities will arise in the construction of $G$ from $g$,
that are usually dealt with by normal ordering. Normal ordering would 
appear to allow us to identify the states $|0\rangle$ and $|0'\rangle$.
However this is not the case, as there are choices of $g$ that are left invariant under 
normal ordering, such as the sum of a holomorphic function plus
an antiholomorphic function, $g(w,\bar w) = g_1(w) + g_2(\bar w)$.
Under such a transformation one can see that the state $|0\rangle$ satisfying 
eqn. (\ref{xvac}) will not satisfy eqn. (\ref{xvacu}). 
We conclude that, in the absence of a complex structure 
on classical phase space, the vacuum depends on the observer. 
The state $|0\rangle$ is only defined locally on ${\cal C}$; it cannot be 
extended globally to all of ${\cal C}$. 

Similar conclusions may be expected for the coherent states $|w^l\rangle$. 
The latter are defined only locally, as eigenvectors of the local annihilation 
operator, with eigenvalues given in equation (\ref{combw}):
\begin{equation}
a^l|w^l\rangle=w^l|w^l\rangle,\qquad l=1,\ldots, n.
\label{annopp}
\end{equation}
Due to $[a^j,a_k^{\dagger}]=\delta^j_k$, under the nonholomorphic 
coordinate transformation (\ref{nonhol}), the local coherent states 
$|w^l\rangle$ are {\it not} mapped into the local coherent states 
satisfying
\begin{equation}
a'^l|w'^l\rangle=w'^l|w'^l\rangle,\qquad l=1,\ldots, n
\label{annoppp}
\end{equation}
in the primed coordinates. No such problems arose for the holomorphic operator 
equation $A'=F(A)$ corresponding to the holomorphic coordinate change $z'=f(z)$ 
of equation (\ref{fhol}), because the commutator $[A^j, A_k^{\dagger}]=\delta^j_k$ 
played no role. Thus coherence becomes a local property on classical phase space.
In particular,  observers not connected by means of a holomorphic change 
of coordinates need not, and in general will not, agree on what is a semiclassical 
effect {\it vs.} what is a strong quantum effect. 

As in section \ref{globcoh}, one can reverse the order of arguments. 
Start from the assumption that, around every point on ${\cal C}$, one can define local 
coherent states $|w^l\rangle$ and a local vacuum $|0\rangle$, that however fall short
of being global. This means that there exists at least one symplectomorphism of ${\cal C}$
that does {\it not}\/ preserve the globality property. Local coordinates $w^l$ around 
any point are defined by the eigenvalue equations (\ref{annopp}). Collecting together 
the set of all such possible local charts we obtain an atlas for ${\cal C}$. 
However, unless the local coherent states $|w^l\rangle$ are actually global, 
this atlas is nonholomorphic. This defines an almost complex structure $J_{\cal C}$.

To summarise, the existence of an almost complex structure $J_{\cal C}$ 
is equivalent to the existence of a locally--defined vacuum and
locally--defined coherent states. When the latter are actually global, then
$J_{\cal C}$ lifts to a complex structure ${\cal J}_{\cal C}$, whose 
associated almost complex structure is $J_{\cal C}$ itself.

\section{Proof of coherence}\label{prov}

We have called {\it coherent} the states constructed in previous sections.
However, we have not verified that they actually satisfy the usual requirements 
imposed on coherent states \cite{COHST}. What ensures that the states so 
constructed are actually coherent is the following argument. 

We have made no reference to coupling constants or potentials, with the understanding 
that the Hamilton--Jacobi method has already placed us, by means of suitable coordinate 
transformations, in a coordinate system on ${\cal C}$ where all interactions vanish. 
At least under the standard notions of classical {\it vs.} quantum, this is certainly 
always possible at the classical level. At the quantum level, the approach of 
ref. \cite{MATONE}, which contains the standard quantum mechanics used here
as a limiting case, rests precisely on the possibility of transforming between any 
two quantum--mechanical states by means of diffeomorphisms. 

Then any dynamical system with $n$ independent degrees of freedom that can be transformed 
into the freely--evolving system can be further mapped into the $n$--dimensional harmonic 
oscillator. The combined transformation is canonical. Moreover it is locally holomorphic when 
${\cal C}$ is an almost complex manifold. Thus locally on ${\cal C}$, our global states 
$|z^l\rangle$ of section \ref{globcoh} coincide with the coherent states of the 
$n$--dimensional harmonic oscillator. Mathematically this fact reflects the structure 
of a complex manifold: locally it is always holomorphically diffeomorphic to 
(an open subset of) ${\bf C}^n$. Physically this fact reflects 
the decomposition into the creation and annihilation modes of perturbative quantum 
mechanics and field theory. In this way, the mathematical problem of patching together different 
local coordinate charts $(U_{\alpha}, z^l_{\alpha})$ labelled by an index 
$\alpha$ may be recast in physical terms. It is the patching together of different 
local expansions into creators $A^{\dagger}_{\alpha}$ and annihilators $A_{\alpha}$, 
for different values of $\alpha$.

In particular, we can write the resolution of unity on ${\cal H}$ 
associated with a holomorphic atlas on ${\cal C}$ consisting of charts
$(U_{\alpha},z^l_{\alpha})$:
\begin{equation}
\sum_{\alpha}\sum_{l=1}^n\int_{\cal C} {\rm d}\mu_{\cal C}\, |z^l_{\alpha}\rangle  \langle 
z^l_{\alpha}|= {\bf 1},
\label{res}
\end{equation}
where ${\rm d}\mu_{\cal C}$ is an appropriate measure (an 
$(n,n)$--differential) on ${\cal C}$. 

Analogous arguments are also applicable to the local states $|w^l\rangle$ 
of section \ref{loccoh}. 
In particular, every almost complex manifold is locally a complex manifold \cite{KN}.
Every holomorphic coordinate chart on ${\cal C}$ is  diffeormorphic to (an open 
subset of) ${\bf C}^n$, so the $|w^l\rangle$ look locally like the coherent 
states of the $n$--dimensional harmonic oscillator.
However, the loss of holomorphicity of ${\cal C}$ 
alters equation (\ref{res}) in one important way. We may write as above
\begin{equation}
\sum_{\alpha}\sum_{l=1}^n\int_{\cal C} {\rm d}\mu_{\cal C}\, |w^l_{\alpha}\rangle  \langle 
w^l_{\alpha}|,
\label{resx}
\end{equation}
but we can no longer equate this to the identity on 
${\cal H}$. The latter is a {\it complex}\/ vector space, while eqn.
(\ref{resx}) allows one at most to expand an arbitrary,
real--analytic function on ${\cal C}$, since the latter is just a 
real--analytic manifold. Hence we cannot equate (\ref{resx})
to ${\bf 1}_{\cal H}$. We can only equate it to the identity on the {\it 
real}\/
Hilbert space of real--analytic functions on ${\cal C}$.
This situation is not new; coherent states without a resolution of unity have been 
analysed in ref. \cite{NOUNITY}, where they have been related to the choice of 
an inadmissible fiducial vector. It is tempting to equate this latter 
choice with the viewpoint advocated here about the vacuum state. 

\section{When are local coherent states also global?}\label{when}

This question can be recast mathematically as follows: when does an almost 
complex structure $J_{\cal C}$ lift to a complex structure ${\cal J}_{\cal C}$? 

\subsection{The Newlander--Nirenberg theorem}\label{nnt}

The almost complex structure $J_{\cal C}$ is said {\it integrable} 
when the Lie bracket $[Z, W]$ of any two holomorphic vector fields $Z$, $W$ 
on ${\cal C}$ is holomorphic. A necessary and sufficient condition for 
$J_{\cal C}$ 
to be integrable is the following. Define the tensor field $N$
\begin{equation}
N(Z,W)=\left[Z,W\right] - \left[J_{\cal C}Z,J_{\cal C}W\right] 
+J_{\cal C}\left[Z, J_{\cal C}W\right] + J_{\cal C}\left[J_{\cal C}Z, W\right].
\label{tors}
\end{equation}
Now the almost complex structure $J_{\cal C}$ lifts to a complex structure 
${\cal J}_{\cal C}$ if and only if the tensor $N$ vanishes identically 
\cite{NN}.

\subsection{Integrable, almost complex structures}\label{cpsm}

When $J_{\cal C}$ is integrable, the set of all holomorphic vector fields defines 
an integrable holomorphic distribution whose integral manifold is ${\cal C}$ 
itself \cite{KN}. A knowledge of this integrable  distribution
of holomorphic vector fields amounts to determining the manifold ${\cal C}$.
Let us see how this comes about. Assume for simplicity that ${\cal C}$ is connected, 
and let $(U_b, \phi_b)$ be a holomorphic chart centred around a basepoint $b\in {\cal C}$. 
Such a holomorphic chart always exists locally. The map 
\begin{equation}
\phi_b:U_b\rightarrow {\bf C}^{n}
\label{homeo}
\end{equation} 
provides local 
holomorphic coordinates around $b$ whose real and imaginary parts can be taken to be 
Darboux coordinates $(q^l, p_l)$, thanks to the assumption of 
compatibility between $J_{\cal C}$ and $\omega_{\cal C}$. 
Let $Z$ be a holomorphic vector field defined 
on $U_b$. We can interpret $Z$ as mapping the chart $(U_{b}, \phi_{b})$ into 
another chart $(U_{Z(b)}, \phi_{Z(b)})$ centred around an infinitesimally 
close basepoint $Z(b)$. Similarly let $W$ be another holomorphic vector field 
mapping $(U_{b}, \phi_{b})$ into $(U_{W(b)}, \phi_{W(b)})$.
Integrability of $J_{\cal C}$ 
means that the Lie bracket $[Z, W]$ maps $(U_{b}, \phi_{b})$
into another holomorphic chart $(U_{[Z,W](b)}, \phi_{[Z,W](b)})$. 
Were $J_{\cal C}$ not integrable, there would exist a holomorphic chart $(U_c, \phi_c)$ centred
around a basepoint $c\in {\cal C}$, and pair of holomorphic vector 
fields $Z, W$ on $U_c$, such that the chart $(U_{[Z,W](c)}, \phi_{[Z,W](c)})$
would not be holomorphic.

Proceeding as described when $J_{\cal C}$ is integrable, we succeed in covering 
${\cal C}$ with a set of holomorphic charts, the transformations between 
them being holomorphic symplectic diffeomorphisms. Physically, 
holomorphicity ensures that the passage from one observer to another 
respects the globality of the notion of coherence and the 
uniqueness of the vacuum. In section \ref{dual} we will relax the complex 
structure ${\cal J}_{\cal C}$ to an almost complex structure $J_{\cal C}$. 
In so doing we will interpret a nonholomorphic mapping such as 
\begin{equation}
(U_c, \phi_c)\rightarrow (U_{[Z,W](c)}, \phi_{[Z,W](c)})
\label{nonc}
\end{equation}
as a duality transformation of the quantum theory.

We can turn things around and recast the Newlander--Nirenberg 
theorem in physical terms: when the commutator of any two (infinitesimal) 
canonical transformations on ${\cal C}$ maps coherent states into coherent states, 
then ${\cal C}$ admits a complex structure. 
The latter is the lift of the almost complex structure $J_{\cal C}$ defined 
by $q^l+{\rm i}p_l$ in terms of Darboux coordinates $q^l$, $p_l$. Conversely, 
if a canonical transformation on ${\cal C}$ maps coherent states into 
noncoherent, or viceversa, then $J_{\cal C}$ does not lift to a complex 
structure.
 
\section{Duality transformations}\label{dual}

When $J_{\cal C}$ is nonintegrable the above construction breaks down.
This gives rise to duality transformations of the quantum theory.

\subsection{Holomorphic foliations of classical phase space}\label{nonint}

Let us consider the case when ${\cal C}$ admits a certain foliation ${\cal F}$ 
by holomorphic, symplectic submanifolds ${\cal L}$ called {\it leaves} \cite{DEWITT}. 
For simplicity we will make a number of technical assumptions. 
First, the leaves ${\cal L}$ have constant real dimension $2m$, 
where $0<2m<2n$; $m$ is called the {\it rank}\/ of the foliation ${\cal F}$. 
We will use the notation $\tilde{\cal L}$ to denote the $2(n-m)$--dimensional 
complement of the ${\cal L}$ in ${\cal C}$. We will assume maximality of 
the rank $m$, {\it i.e.}, no holomorphic leaf exists with dimension 
greater than $2m$. Second, we suppose that the restrictions 
$\omega_{\cal L}$  and $\omega_{\tilde {\cal L}}$ of the symplectic form $\omega_{\cal C}$ 
render the leaves ${\cal L}$ and their complements $\tilde{\cal L}$ symplectic.
Third we assume that, on the ${\cal L}$, the complex structure is 
compatible with the symplectic structure as in section \ref{globcoh}. 
Fourth, the complement $\tilde {\cal L}$ is also assumed to carry an almost 
complex structure compatible with $\omega_{\tilde {\cal L}}$ as in section \ref{loccoh}.

All these assumptions amount to a decomposition of $\omega_{\cal C}$
as a sum of two terms,  
\begin{equation}
\omega_{\cal C}=\omega_{\cal L} + \omega_{\tilde {\cal L}},
\label{twot}
\end{equation}
where in local Darboux coordinates around a basepoint $b\in {\cal C}$ we have
\begin{equation}
\omega_{\cal L}=
\sum_{k=1}^{m} {\rm d}p_k\wedge {\rm d}q^k, \qquad 
\omega_{\tilde {\cal L}}=
\sum_{j=m+1}^{n} {\rm d}p_j\wedge {\rm d}q^j.
\label{twoterms}
\end{equation}
Furthermore the combinations $z^k=q^k+{\rm i}p_k$, $k=1,\ldots, m$, 
are holomorphic coordinates on the ${\cal L}$, while the combinations 
$w^j=q^j+ {\rm i} p_j$, $j=m+1, \ldots, n$, are coordinates on $\tilde{\cal L}$. 
In this way a set of coordinates around $b$ is
\begin{equation}
z^1, \ldots, z^{m}, w^{m+1}, \ldots, w^n.
\label{quacoords}
\end{equation}
The holomorphic leaf ${\cal L}$ passing through $b$ may be taken to be
determined by
\begin{equation}
w^{m+1}=0,\ldots, w^n = 0,
\label{pass}
\end{equation}
and spanned by the remaining coordinates $z^k$, $k=1,\ldots, m$.

The construction of the previous sections can be applied as follows. 
Coherent states $|z^k;w^j\rangle$ can be defined locally on 
${\cal C}$. They cannot be extended globally over all of ${\cal C}$, 
as the latter is not a complex manifold. However the foliation ${\cal F}$ 
consists of holomorphic submanifolds ${\cal L}$. On each one of them there 
exist global coherent states specified by equations 
(\ref{quacoords}), (\ref{pass}), {\it i.e.}, 
\begin{equation}
|z^k;w^{m+1}=0,\ldots, w^n = 0\rangle.
\label{speci}
\end{equation}
Physically, this case corresponds to a fixed splitting of the $n$ degrees 
of freedom in such a way that the first $m$ of them give rise to global coherent 
states on the holomorphic leaves ${\cal L}$. On the latter there is no room 
for nontrivial dualities. On the contrary, the last $n-m$ degrees of freedom
are only locally holomorphic on ${\cal C}$. Holomorphicity is lost 
globally on ${\cal C}$, thus allowing for the possibility of nontrivial 
duality transformations between different holomorphic leaves ${\cal L}$. 

Let us analyse the resolution of unity in terms of the states
$|z^k;w^j\rangle$. With the notations of section \ref{prov}, the expansion
\begin{equation}
\sum_{\alpha}\sum_{k=1}^m\sum_{j=m+1}^n\int_{\cal C} {\rm d}\mu_{\cal C}\, 
|z^k_{\alpha};w^j_{\alpha}\rangle  \langle w^j_{\alpha};z^k_{\alpha}|
\label{newres}
\end{equation}
cannot be equated to the identity,
for the same reasons as in section \ref{prov}.
However, integrating over the $w^l$, the expansion
\begin{equation}
\sum_{\alpha}\sum_{k=1}^m\int_{\cal L} {\rm d}\mu_{\cal L}\, 
|z^k_{\alpha}\rangle  \langle z^k_{\alpha}|
\label{newresx}
\end{equation}
can be equated to the identity. The integral extends over any one leaf ${\cal L}$ 
of the foliation ${\cal F}$. On the contrary, integrating over the $z^l$ 
in (\ref{newres}) would not give a resolution of the identity.

\subsection{Examples of duality groups}\label{otros}

The previous sections illustrate a possible mechanism to realise quantum--mecha-nical 
duality transformations between different vacua and between the coherent states built 
around them. Holomorphic foliations ${\cal F}$ such as those of section \ref{nonint} 
allow for both continuous and discrete duality transformations. Assume that the $w^l$ span 
a real, $2(n-m)$--dimensional manifold invariant under a certain group ${\cal D}$ 
of nonholomorphic transformations. Then ${\cal D}$ becomes a duality group 
of the quantum theory on ${\cal Q}$. In principle, appropriate choices of the 
holomorphic foliations ${\cal F}$ will allow to obtain any given duality group. 

A simple example of a holomorphic foliation (that also happens to be 
a symplectic foliation of a Poisson manifold) is given by the Kirillov form 
\cite{LIBAZCA} for the Lie algebra su(2).  Using coordinates $x$, $y$, $z$, the latter
is spanned by generators $T_x$, $T_y$, $T_z$ satisfying the commutation relations
\begin{equation}
\left[ T_i, T_j\right]=\epsilon_{ijk}T_k.
\label{commut}
\end{equation}
There is the Casimir operator $T_x^2+T_y^2+T_z^2$ on the enveloping algebra 
of su(2). We have a Poisson tensor $P$
\begin{equation}
P=\left( \begin{array}{ccc}
 0& z& -y\\
-z& 0& x\\
 y& -x& 0
\end{array} \right),
\label{matriz}
\end{equation}
and the Poisson bracket of any two functions $f$, $g$ on su(2) reads
$$
\left\{f,g\right\}= 
x\left(\partial_y f \partial_z g - \partial_z f \partial_y g\right) 
$$
\begin{equation}
-y\left(\partial_x f \partial_z g - \partial_z f \partial_x g\right) +
z\left(\partial_x f \partial_y g - \partial_y f \partial_x g\right).
\label{poibra}
\end{equation}
Now ${\rm det} P=0$ everywhere. Away from the origin $x=y=z=0$, the Poisson matrix $P$
always contains a $2\times 2$ nonsingular submatrix; this corresponds to 
the existence of the Casimir function $f(x,y,z)=x^2+y^2+z^2$. Hence we have 
a symplectic foliation of su(2) by symplectic leaves which are concentric spheres 
$x^2+y^2+z^2=R^2$ of increasing radii $R>0$. Passing through each point of su(2) 
there is exactly one such symplectic leaf; only the origin is met by no leaf, 
as the foliation has  zero rank there. In standard spherical coordinates 
$r, \theta, \varphi$  one 
finds a Poisson bracket on the leaves 
\begin{equation}
\left\{f,g\right\}={1\over r\sin\theta}\left(\partial_{\theta}f 
\partial_{\varphi}g - \partial_{\varphi}f \partial_{\theta}g\right)
\label{ssphc}
\end{equation}
and a Kirillov symplectic form 
\begin{equation}
K=r\sin \theta\, {\rm d} \varphi\wedge {\rm d} \theta.
\label{kiri}
\end{equation}

To us, the above example is interesting not because the leaves are symplectic, 
but because they are holomorphic. They provide a holomorphic foliation of 
su(2). Of course, the latter cannot be a symplectic manifold, 
but one can imagine embedding these holomorphic leaves into a nonholomorphic, 
symplectic manifold.

\section{Discussion}\label{discussion}

\subsection{Recapitulation}\label{recap}

Most physical systems admit a complex structure ${\cal J}_{\cal C}$ on their 
classical phase spaces ${\cal C}$. Prominent among them is the 
1--dimensional harmonic oscillator. Mathematically, the corresponding 
${\cal C}$ supports the simplest holomorphic structure, that of the complex plane.
Physically, canonical quantisation rests on the decomposition 
of a field into an infinite number of oscillators. The notions that the vacuum state 
is unique, and that coherence is a universal property independent of the 
observer, follow naturally. However, as summarised in section \ref{intro}, 
recent breakthroughs in quantum field theory and M--theory suggest the need 
for a framework in which duality transformations can be accommodated at the elementary 
level of quantum mechanics, before considering field theory or strings. 
This, in turn, would help to understand better the dualities underlying 
quantum fields, strings and branes.

The formalism presented here can accommodate duality transformations in a 
natural way. In the absence of a complex structure ${\cal J}_{\cal C}$, 
all our statements concerning the vacuum state and the property of coherence are 
necessarily local in nature, {\it i.e.}, they do not hold globally on ${\cal C}$. 
A duality transformation of the quantum theory on ${\cal Q}$ will thus be specified 
by a nonholomorphic coordinate transformation on ${\cal C}$.

However, the question immediately arises: do we not have an overabundance 
of vacua? Does every imaginable nonholomorphic transformation induce a 
{\it physical}\/ duality? A judicious application of physical symmetries 
can vastly restrict this apparent overabundance. Usually dualities appear under the form 
of a group ${\cal D}$. Rather than taking every imaginable nonholomorphic transformation 
to define a physical duality we must assume, as is the case in M--theory, 
a knowledge of the duality group ${\cal D}$, or perhaps even a finite subgroup thereof, 
and restrict ourselves to those nonholomorphic transformations that actually realise it. 

\subsection{The notion of duality}\label{notdual}

Duality is not to be understood as a transformation between different physical 
phenomena. Rather, it is to be understood as a transformation between different 
descriptions of the same quantum physics. Similarly, the statement that $\hbar$ 
depends on the observer is to be understood as meaning that one given quantum 
phenomenon may be described by different observers on ${\cal C}$ as corresponding 
to different regimes in a series expansion in powers of $\hbar$. Thus
the semiclassical regime is given by a truncation of this series to order $\hbar$, 
while the strong quantum regime requires the whole infinite expansion.

That coherence equals holomorphicity has been known for long \cite{LONG}.
Here we have proved that noncomplex structures (such as almost 
complex structures) allow to implement duality transformations.
The picture that emerges is conceptually close to that of a 
geometric manifold covered by local coordinate charts. Instead of the 
latter, quantum mechanics has local coherent states. A change of Darboux
coordinates on ${\cal C}$ may or may not be holomorphic. Correspondingly, 
a transformation between quantum--mechanical states may or may not preserve 
coherence. Those that do not preserve coherence are duality transformations. 
A duality appears when it is possible to give two or more, apparently different, 
descriptions of the same quantum--mechanical phenomenon. Coherence thus becomes 
a local property on classical phase space ${\cal C}$. Observers on ${\cal C}$ not 
connected by means of a holomorphic change of coordinates need not, and in general 
will not, agree on what is a semiclassical effect {\it vs.} what is a strong quantum 
effect. 

\subsection{Holomorphic foliations  {\it vs.} symplectic foliations}
\label{poistruct}

Some authors take the view that classical phase space ${\cal C}$ requires no more 
structure than a Poisson bracket as the latter becomes, under quantisation, 
the commutator of quantum operators \cite{DIRAC}. A more geometric 
viewpoint \cite{ARNOLD, GQUANT} is to take ${\cal C}$ to be not just Poisson, 
but symplectic. This is also useful in analysing physical issues such as constrained 
dynamics \cite{JACKIW}. 

In implementing duality transformations we have relaxed the complex structure 
${\cal J}_{\cal C}$ to an almost complex structure $J_{\cal C}$. In foliating ${\cal C}$ 
by holomorphic leaves, our approach is reminiscent of the symplectic foliations 
of Poisson manifolds \cite{WEINSTEIN}. Our conclusion is that, just as symplectic 
foliations of Poisson manifolds implement constraints, holomorphic foliations of  
symplectic manifolds implement dualities. 

In our analysis, the symplectic structure of classical phase space ${\cal C}$ 
plays a key role. Moreover, we claim that $\omega_{\cal C}$ also has a quantum--mechanical 
role to play, too. In ref. \cite{STATEMENT} we have  put forward a starting point for a 
formulation of quantum mechanics that is compatible with the relativity of $\hbar$. 
This approach is based on the symplectic structure $\omega_{\cal C}$. In 
fact Darboux's theorem falls short (by $\hbar$) of being a quantisation, as
symplectic geometry does not know about Planck's constant $\hbar$. 
Once supplemented with the physical input $\hbar$, Darboux's theorem truly becomes 
a quantisation. 

\subsection{Final comments}\label{comm}

In using coherent states our approach has been geometric. 
Indeed coherent states have been applied to geometric quantisation 
\cite{GQCS}; closely related issues have been studied recently in 
refs. \cite{HALL, ZA, FUJII, TORONTO}.
However there are a number of alternative viewpoints in order
to analyse duality in quantum mechanics. Planck's constant $\hbar$ 
can be interpreted as the only modulus existing in quantum mechanics. 
It is precisely this parameter that tells {\it classical}\/ from {\it quantum}, 
so duality in quantum mechanics necessarily refers to $\hbar$.  
This interpretation is especially natural in deformation quantisation 
\cite{BEREZIN, SCHLICHENMAIER, GARCIA}. It has gained renewed interest 
in the physical community due to its links with noncommutative theories, 
nontrivial Neveu--Schwarz $B$--fields and branes. It would be very interesting 
to analyse duality from this perspective. 

Along other lines, interesting points to explore in this context are the 
higher--order generalisations of Poisson structures \cite{AZCA} and 
supersymmetric quantum mechanics \cite{RRAGA}. Of course, quantum gravity is 
always an important testing ground for geometric theories \cite{JANAN, THIEMANN, WAT, RAD}. 
Inasmuch as we are relativising the concept of a quantum we may be said
to be quantising gravity too---only in reverse.

{\bf Acknowledgements}

Support from PPARC (grant PPA/G/O/2000/00469) and DGICYT (grant PB 96-0756) is acknowledged.


\begin{thebibliography}{99}

\bibitem{COHST}   
J. Klauder and B.--S. Skagerstam, {\it Coherent States}, World Scientific, Singapore (1985);
A. Perelomov, {\it Generelized Coherent States and their Applications}, Springer Texts and
Monographs in Physics, Berlin (1986). For recent reviews see J. Klauder, {\tt quant-ph/9810043};
J. Klauder, {\tt quant-ph/0110108}.

\bibitem{DUALITY}
D. Olive and P. West (eds.), {\it Duality and Supersymmetric Theories}, Cambridge 
University Press, Cambridge (1999);
M. Kaku, {\it Strings, Conformal Fields and M--Theory},  Springer, Berlin (2000).

\bibitem{LAG}
L. Alvarez--Gaum\'e and S. Hassan, {\it Fortsch. Phys.} {\bf 45} (1997) 159.

\bibitem{VAFA}
C. Vafa, {\tt hep-th/9702201}.

\bibitem{GALINDO}
A. Galindo and P. Pascual, {\it Quantum Mechanics}, vols. I, II, Springer, Berlin 
(1990).

\bibitem{ASHTEKAR}
A. Ashtekar and T. Schilling, {\it Geometrical Formulation of Quantum 
Mechanics}, in {\it On Einstein's Path}, A. Harvey (ed.), Springer, 
Berlin (1999);
A. Ashtekar, {\tt qr-qc/9901023}.

\bibitem{ME}
J.M. Isidro, {\tt hep-th/0110151}.

\bibitem{ANANDAN}
J. Anandan and Y. Aharonov, {\it Phys. Rev. Lett.} {\bf 65} (1990) 1697;
J. Anandan, {\it Found. Phys.} {\bf 21} (1991) 1265.

\bibitem{MATONE}  
A. Faraggi and M. Matone,
{\it Phys. Rev. Lett.} {\bf 78} (1997) 163;
{\it Phys. Lett.} {\bf A249} (1998) 180; 
{\it Phys. Lett.} {\bf B437} (1998) 369;
{\it Phys. Lett.} {\bf B445} (1998) 77;
{\it Phys. Lett.} {\bf B445} (1998) 357;
{\it Phys. Lett.} {\bf B450} (1999) 34;
{\it Int. J. Mod. Phys.} {\bf A15} (2000) 1869;
G. Bertoldi, A. Faraggi and M. Matone, 
{\it Class. Quant. Grav.} {\bf 17} (2000) 3965.

\bibitem{FH}
L. Hughston and D. Brody, {\it Proc. Roy. Soc.} {\bf A454} (1998) 2445;
{\it J. Math. Phys.} {\bf 39} (1998) 6502;
L. Hughston and T. Field, {\it J. Math. Phys.} {\bf 40} (1999) 2568.

\bibitem{GEOMKLAUDER}
J. Klauder, {\tt quant-ph/0112010}.

\bibitem{MILANO}
R. Cirelli, M. Gatti and A. Mani\`a, {\tt quant-ph/0202076}.

\bibitem{TIWARI}
S. Tiwari, {\tt quant-ph/0109048}.

\bibitem{MEX}
H. Garc\'{\i}a--Compe\'an, J. Pleba\'nski, M. Przanowski and F. 
Turrubiates, {\tt hep-th/0112049};
I. Carrillo--Ibarra and H. Garc\'{\i}a--Compe\'an, {\tt hep-th/0202015}.

\bibitem{MCDUFF}
D. McDuff and D. Salamon, {\it Introduction to Symplectic Topology},
Oxford University Press, Oxford (1998);
D. McDuff, {\tt math.SG/0201032}.

\bibitem{ARNOLD}
V. Arnold, {\it Mathematical Methods of Classical Mechanics}, Springer, Berlin (1989).

\bibitem{KN}
S. Kobayashi and K. Nomizu, {\it Foundations of Differential Geometry}, Wiley, 
New York (1996).

\bibitem{GQUANT}
\`Sniatycki, {\it Geometric Quantization and Quantum Mechanics}, Springer, Berlin (1980);
N. Woodhouse, {\it Geometric Quantization}, Oxford University Press, Oxford (1991).

\bibitem{BEREZIN} 
F. Berezin, {\it Sov. Math. Izv.} {\bf 38} (1974) 1116; 
{\it Sov. Math. Izv.} {\bf 39} (1975) 363;
{\it Comm. Math. Phys.} {\bf 40} (1975) 153;
{\it Comm. Math. Phys.} {\bf 63} (1978) 131.

\bibitem{SCHLICHENMAIER}
M. Schlichenmaier, {\it Berezin--Toeplitz Quantization and Berezin's 
Symbols for Arbitrary Compact K\"ahler Manifolds}, in {\it Coherent 
States, Quantization and Gravity}, M. Schlichenmaier {\it et al.} (eds.),
Polish Scientific Publishers PWN, Warsaw (2001);
S. Berceanu and M. Schlichenmaier, {\tt math.DG/9903105};
M. Schlichenmaier, {\tt math.QA/9910137}, {\tt math.QA/0005288}; 
A. Karabegov and M. Schlichenmaier, {\tt math.QA/0006063}, {\tt  
math.QA/0102169}.

\bibitem{GARCIA}
H. Garc\'{\i}a--Compe\'an, J. Pleba\'nski, M. Przanowski and F. 
Turrubiates, {\it Int. J. Mod. Phys.} {\bf A16} (2001) 2533.

\bibitem{NOUNITY}
J. Klauder, {\tt quant-ph/0008132}.

\bibitem{NN}
A. Newlander and L. Nirenberg, {\it Ann. Math.} {\bf 65} (1957) 391;
A Nijenhuis and W. Woolf, {\it Ann. Math.} {\bf 77} (1963) 424.

\bibitem{DEWITT}
Y. Choquet--Bruhat and C. DeWitt--Morette, {\it Analysis, Manifolds and 
Physics}, vols. I, II, North--Holland, Amsterdam (1989).

\bibitem{LIBAZCA}
J. de Azc\'arraga and J. Izquierdo, {\it Lie Groups, Lie Algebras, 
Cohomology and some Applications in Physics}, Cambridge University Press, 
Cambridge (1995).

\bibitem{LONG}
J. Klauder, {\it Ann. Phys.} {\bf 11} (1960) 123;
V. Bargman, {\it Comm. Pure. Appl. Math.} {\bf 14} (1961) 187.

\bibitem{DIRAC}
P. Dirac, {\it Lectures on Quantum Mechanics}, Yeshiva Press, New York (1964);
{\it The Principles of Quantum Mechanics}, Oxford University Press, Oxford (2001).

\bibitem{JACKIW}
L. Faddeev and R. Jackiw, {\it Phys. Rev. Lett.} {\bf 60} (1988) 1692;
R. Jackiw, {\it Constrained Quantization without Tears}, in {\it Constraint 
Theory and Quantization Methods}, F. Colmo {\it et al.} (eds.), World 
Scientific, Singapore (1994); R. Jackiw, {\it Diverse Topics in 
Theoretical and Mathematical Physics}, World Scientific, Singapore (1995).

\bibitem{WEINSTEIN}
A. Weinstein, {\it J. Diff. Geom.} {\bf 18} (1983) 523.

\bibitem{STATEMENT}
J.M. Isidro, {\tt quant-ph/0112032}.

\bibitem{GQCS}
J. Klauder, {\tt quant-ph/9510008}.

\bibitem{HALL}
B. Hall, {\tt quant-ph/0012105};
B. Hall and J. Mitchell, {\tt quant-ph/0109086}, {\tt quant-ph/0203142}.

\bibitem{ZA}
L. Boya, A. Perelomov and M. Santander, {\tt math-ph/0111022}.

\bibitem{FUJII}
K. Fujii, {\tt quant-ph/0112090}, {\tt quant-ph/0202081}.

\bibitem{TORONTO}
S. Bartlett, D. Rowe and J. Repka, {\tt quant-ph/0201129, 
quant-ph/0201230}.

\bibitem{AZCA}
J. de Azc\'arraga, A. Perelomov and J. P\'erez Bueno, {\it J. Phys. A} 
{\bf 29} (1996) 7993;
J. de Azc\'arraga, A. Perelomov and J. P\'erez Bueno, {\it J. Phys. A: 
Math. Gen.} {\bf 29} (1996) L151;
J. de Azc\'arraga, J. Izquierdo and J. P\'erez Bueno, {\it J. Phys. A: 
Math. Gen.} {\bf 30} (1997) L607.

\bibitem{RRAGA}
J. de Azc\'arraga, J. Izquierdo and A. Macfarlane, {\it Nucl. Phys.} 
{\bf B604} (2001) 75.

\bibitem{JANAN}
J. Anandan, {\tt quant-ph/0012011}.

\bibitem{THIEMANN}
T. Thiemann, {\tt gr-qc/0110034}.

\bibitem{WAT}
G. Watson and J. Klauder, {\tt gr-qc/0112053}.

\bibitem{RAD}
For recent reviews see A. Ashtekar, {\tt gr-qc/0112038}, {\tt math-ph/0202008}.

\end{thebibliography}
\end{document}